\documentstyle[PASJadd]{PASJ95}
\newcommand{\vol}{51}
\newcommand{\no}{4}
\newcommand{\lett}{}
\newcommand{\spage}{}
\newcommand{\rdate}{March 5, 1999}
\newcommand{\adate}{July 5, 1999}

\def\etal{et al.\ }

\def\apj    {ApJ{\rm}\ }
\def\apjs   {ApJS{\rm}\ }
\def\aj     {AJ{\rm}\ }
\def\aap    {A{\rm\&}A{\rm}\ }

\def\nat    {Nature{\rm}\ }

\def\sci    {Science{\rm}\ }

\begin{document}

\title{High Resolution VSOP Imaging of 
a Southern Blazar PKS~1921--293 at 1.6~GHz}

\author{Z.-Q.\ {\sc Shen},$^{1,3}$ 
P.\ G.\ {\sc Edwards},$^2$ J.\ E.\ J.\ {\sc Lovell},$^2$
K.\ {\sc Fujisawa},$^1$ S.\ {\sc Kameno}, $^1$ and M.\ {\sc Inoue}$^1$ \\[12pt]
$^1${\it National Astronomical Observatory, 2-21-1 Osawa, Mitaka, Tokyo 181-8588} \\
$^2${\it Institute of Space and Astronautical Science, 3-1-1 Yoshinodai, 
Sagamihara, Kanagawa 229-8510} \\
$^3${\it Academia Sinica Institute of Astronomy and Astrophysics, 
P.O. Box 1-87, Nankang, Taipei 115} \\ 
{\it E-mail(ZS): zshen@hotaka.mtk.nao.ac.jp}}

\abst{We present a high resolution 1.6~GHz VSOP image of the southern blazar 
PKS~1921--293. The image shows a typical core--jet morphology, consistent with 
ground--based VLBI images. However, the addition of data from the space antenna 
has greatly improved the angular resolution (especially along the north--south 
direction for this source), and thus allowed us to clearly identify the core. 
Model fitting reveals an inner jet component $\sim$1.5~mas north of the core.
This jet feature may be moving on a common curved path connecting the 
jet within a few parsecs
 to the 10--parsec--scale jet.  
The compact core has a brightness temperature of 2.6~$\times$~10$^{12}$~K 
(in the rest frame of the quasar), an indication of relativistic beaming. 
We analyzed the source in terms of three models, involving the inverse Compton 
catastrophe, an inhomogeneous relativistic jet, and the equipartition of energy
between the radiating particles and the magnetic field. 
Our analysis of this $\gamma-$ray--quiet blazar shows no preference to any
particular one of these models.
} 

\kword{Galaxies: active --- Galaxies: nuclei --- Quasars: individual (PKS~1921--293)}

\maketitle
\thispagestyle{headings}

\section{Introduction}

The successful launch of the VLBI Space Observatory Programme (VSOP) satellite
HALCA marks a great step forward in increasing the resolution over that 
possible with ground-based radio telescopes at 1.6 and 5.0~GHz 
(Hirabayashi \etal 1998 and references therein). 
HALCA's 8-meter-diameter antenna is in an elliptical orbit with an 
apogee of 21,400~km, a perigee of 560~km and an orbital period of 6.3~hours. 
VSOP observations, with a factor of $\sim$3 improvement in resolution 
compared to the ground observations at same frequencies, 
enable a close look at the compact core of active galactic nuclei and as a 
result, to resolve some individual components within the compact core and jets, 
and to study the bent jet in the vicinity of the core observed at higher 
frequencies with ground telescopes. 
%The full sky coverage of the satellite's 
%orbit with proper scheduling will allow a good {\sl u-v} sampling for imaging. 
In particular, the addition of HALCA significantly improves the north--south 
resolution for equatorial and southern radio sources, as illustrated in this paper.
VSOP also provides almost an order of 
magnitude increase to the detectable brightness temperature (from 
10$^{11}$--10$^{12}$~K to 10$^{12}$--10$^{13}$~K for bright sources).

As a highly polarized (cf. Worrall, Wilkes 1990) and optically violently 
variable quasar (Wills, Wills 1981) with m$_v$~=~17.5, PKS~1921--293 (OV--236)
is classified as one of the brightest radio--loud blazars known.
It shows a dramatic variability from radio to X-ray. 
Curiously, no $\gamma$--ray emission has been detected by EGRET (Fichtel 
\etal 1994; Mukherjee \etal 1997). 
At a redshift of 0.352 (Wills, Wills 1981), it has an 
angular--to--linear scale conversion of 3$h^{-1}$~pc~mas$^{-1}$ with 
H$_0$~=~100~$h$~km~s$^{-1}$~Mpc$^{-1}$ and q$_0$~=~0.5. 
The existing ground VLBI observations reveal a core--jet structure 
(cf. Kellermann \etal 1998).
Its core is very compact (only a fraction of the beamwidth in diameter), with
a brightness temperature (T$_b$) in the rest frame of the source greater 
than 10$^{12}$~K. 
There is evidence that on a scale of 1--2~$h^{-1}$~pc from the core, the jet 
moves along a curved trajectory superluminally 
(Shen \etal 1999) and then appears to end up in a diffuse component about 
15~$h^{-1}$~pc from the core (cf. Tingay \etal 1998).

In this letter, we report on the results of a 1.6~GHz VSOP imaging of
PKS~1921--293. We describe the observations and data reduction and present a 
1.6~GHz VSOP image of PKS~1921--293 in section 2. The evolution of its fine
structure and the implication of the high T$_b$ will be discussed in section 
3. A brief summary is given in section 4.

Throughout this paper the spectral index, $\alpha$, is defined as 
S$_\nu$~$\propto$~$\nu^\alpha$.

\section{VSOP Observations and Data Reduction}

The 1.6~GHz VSOP observations of PKS~1921--293 were carried out as part of 
HALCA's in--orbit checkout on July 18, 1997 for a total about 1.5 hours.
The HALCA data acquisition was successfully done with the satellite
tracking stations located in Goldstone (CA, USA), and NRAO 
\footnote {The NRAO is operated by Associated Universities Inc., under 
cooperative agreement with the National Science Foundation} 
Green Bank (WV, USA). 
The ground radio telescopes consisted of 10 VLBA antennas and 
the phased VLA of NRAO. 
The left-circular polarization (LCP) data were recorded in the standard VLBA 
format with an intermediate frequency (IF) band of 16~MHz.
The cross-correlation of the data was carried out on the VLBA correlator at 
Socorro (NM, USA) with an output preaveraging time of 0.524 and 1.966 seconds 
for the space-ground and ground-ground baselines, respectively, and 256
spectral channels per IF band.

The post--correlation data reduction was performed in NRAO AIPS and 
DIFMAP (Shepherd 1997). 
{\sl A priori} visibility amplitude calibrations were applied using the 
antenna gain curves and the system temperatures measured at each antenna 
including HALCA. 
In the fringe--fitting run, a solution interval of 1 minute and a point 
source model were employed. 
The VLBA antenna at Los Alamos (LA) served as the reference telescope 
throughout. 
Strong fringes were consistently detected on space baselines to HALCA as well 
as all the ground baselines. 
Following this, the data were averaged over all frequency channels, and then 
phase self-calibrated with a 10--second solution interval and a point source 
model for the purpose of further time averaging. 

Finally, the visibility data were exported to DIFMAP for imaging.
The data were integrated over 30 seconds to reconcile the different 
preaveraging time from the correlator output as mentioned above. 
The uncertainties in the averaged visibilities were computed from the scatter
of data points within the averaging interval. 
Some obviously bad data were inspected and removed.
Several iterations of cleaning and self--calibration to phases (and amplitudes
in the later stages) were performed. 
To ensure a better angular resolution with HALCA data, uniform weighting of 
the data was adopted with gridding weights scaled by amplitude errors raised 
to the power of --1. 
The resulting image is shown in figure~1. 
The FWHM beam size is 4.1~mas~$\times$~1.1~mas at a position angle of 46$^\circ$. 
(For comparison, the synthesized beam of the ground--only observation 
is 21.7~mas~$\times$~5.7~mas along --4$^\circ$.) 
The peak flux density and the rms noise level are 4.61~Jy/beam and 7.0~mJy/beam,
respectively. 
Thus, a peak--to--rms dynamic range of 650 is obtained in our short 1.6~GHz VSOP 
image. 

\begin{fv}{1}{18pc}
{A 1.6~GHz VSOP image of PKS~1921--293 observed at epoch 1997.55
(July 18, 1997) with VLBA and the phased VLA. The restoring beam is 
4.1~mas~$\times$~1.1~mas at a position angle of 46$^\circ$ (indicated at the lower 
left corner). Contour levels are drawn at --35, 35~$\times$~1.8$^n$~mJy~beam$^{-1}$
(n~=~0, ..., 8), and the peak flux density is 4.61~Jy~beam$^{-1}$.}
\end{fv}

\begin{fv}{2}{18pc}
{A plot shows the model fitting (solid curves) to the visibility amplitude as a 
function of {\sl u-v} distance with a plot of {\sl u-v} coverage embedded.}
\end{fv}

\section{Discussion}

\begin{table*}
\small
\begin{center}
Table~1.  \hspace{4pt} Results from Model Fitting to 1.6~GHz VSOP Observation \\
\end{center}
\vspace{6pt}
\begin{tabular*}{\textwidth}{@{\hspace{\tabcolsep}
\extracolsep{\fill}}p{7pc}ccccccc}
\hline\hline \\ [-6pt]
Component & S &  r & $\theta$ & a & b/a & P.A. \\ 
     & (Jy) & (mas)  & ($^\circ$) &  (mas)  &   &  ($^\circ$) \\ [4pt]\hline\\[-6pt]
1 \dotfill & 5.59$\pm$0.11 &      0        &     0        & 1.87$\pm$0.15 
& 0.38$\pm$0.03 & 48.9$\pm$3.4   \\
2 \dotfill & 2.03$\pm$0.05 & 1.5$\pm$0.2 & 1.3$\pm$4.0  & 2.70$\pm$0.12 
& 0.12$\pm$0.01 & 48.0$\pm$3.4   \\
3 \dotfill & 4.63$\pm$0.10 & 5.6$\pm$0.1 & 30.4$\pm$1.3 & 3.52$\pm$0.14 
& 0.76$\pm$0.08 & 117.8$\pm$1.7  \\ [4pt]
\hline
\end{tabular*} 
\vspace{6pt} \par\noindent
{\small Notes $-$ { 
S: the flux density of each component; (r, $\theta$): the distance and position angle 
of each component with respect to the origin defined by component 1 in mas and degrees,
respectively; (a, b/a and P.A.): three parameters of Gaussian component, i.e. major
axis (FWHM) in mas, ratio of minor to major axes and the orientation angle in degrees 
of the major axis}}
\end{table*}

\subsection{Structural Evolution}

PKS~1921--293 was unresolved at arcsecond--scale with VLA observations (Perley 1982;
de Pater \etal 1985). 
Ground VLBI images at centimeter wavelengths showed a typical core--jet structure, 
with a diffuse jet feature located at a position angle $\sim$~30$^\circ$ with respect 
to the compact, strong core (cf. Fey \etal 1996; Tingay \etal 1998;
Kellermann \etal 1998). 
At 43~GHz, three--epoch VLBA images provide evidence for a superluminal jet 
($\beta_{app}$~=2.1~$h^{-1}$) within 1--2~$h^{-1}$~pc, which has a sharp bend in 
the position angle compared to the jet seen on a scale of ten parsecs 
(Shen \etal 1999). 

The core--jet morphology of our VSOP image is in good agreement with those 
ground VLBI images made at other centimeter wavelengths. 
However, the addition of space VLBI antenna greatly improved the resolution 
(as can be seen from the comparison of the beams with and without HALCA), and 
thus enables us to clearly identify the compact core. 
To yield a quantitative description of the source structure, we applied a model 
consisting of three elliptical Gaussian components to fit both amplitudes and 
phases in the calibrated visibility data. 
The results of model parameters and corresponding 1-$\sigma$ errors 
are listed in table~1. 
It reveals that the data are consistent with an inner jet (component~2) 
at 1.5~mas north of the core (component~1), as well as a large jet feature 
(component~3) at a position angle of 30$^\circ$\hspace{-4.5pt}.\hspace{.5pt}4 
and a separation of 5.6~mas from the core. 
We note that the orientation angles of the two central components are practically 
identical, and both are within errors the same as the position angle of the 
synthesized beam for this observation.
We do not regard these coincidences as being physically meaningful, although the 
model--fitting procedure gives the best results with these values. 
We are confident, however, that the relative separation and position angle of the 
components are correct.

We show in figure~2 the distribution of visibility amplitude versus {\sl u-v} 
distance along with the visibilities generated from our model fit (solid curves). 
The plot shows a dramatic drop of the correlated flux density within the ground 
baselines ($<$~30~M$\lambda$), associated with the resolved diffuse jet 
(component~3) seen in all centimeter ground VLBI images. 
On the space baselines ($>$~50~M$\lambda$), there is still a clear but gradual 
decrease in amplitude (to $\sim$~1.0~Jy on the longest baselines of 
140~M$\lambda$), which indicates that the compact central part is no longer 
unresolved.
The measured non--zero closure phases on the HALCA baselines also confirm this.
The introduction of component~2 is required to fit the visibility distribution
at a {\sl u-v} distance range of (50~--~70)~M$\lambda$, which is eventually 
composed of space baselines in the north-south direction.
We also tried additional model fitting to visibilities on the space baselines only. 
This should fit compact core structure well since the extended 
structure at about 5.6~mas away is totally resolved. 
The fit gives a separation of two components about 1.6~mas along  
3$^\circ$\hspace{-4.5pt}.\hspace{.5pt}3,
consistent with the results in table~1. 
We note that component~2 is spatially coincident with an observed 43~GHz weak 
extended feature of $\sim$18~mJy (Shen \etal 1999).
If these are the same component at the two frequencies, then the spectral index
between 1.6 and 43~GHz would be steeper than --1.6, which is very common in the
optically thin jets.
However, since PKS~1921--293 is highly variable, this estimate 
of spectral index made from measurements taken at different epochs 
(1.5 years apart) may not be accurate and, further observations are
definitely required to clarify this.
We found that the position of this inner northern jet component from our 
1.6~GHz VSOP experiment with 7 times better resolution in the north-south 
direction than ground-only VLBI observations (see {\sl u-v} coverage embedded 
in figure~2), when compared with those ground VLBI images, is located on a 
common curved path connecting the jet within 1--2~$h^{-1}$~pc to the 10~pc--scale jet. 

\subsection{Brightness Temperature T$_b$}

PKS~1921--293 has one of the highest brightness temperatures measured in the
rest frame of the source. 
A 22~GHz VLBI survey (Moellenbrock \etal 1996) gave a lower limit to 
T$_b$~$>$~7.0$^{+4.0}_{-2.1}$~$\times$~10$^{12}$~K for PKS~1921--293.
A previous VLBI experiment, using a telescope in Earth orbit, estimated a core 
T$_b$ of 3.8~$\times$~10$^{12}$~K at 2.3~GHz (Linfield \etal 1989), the highest 
in the sample for sources with known redshifts. 
VLBI images made at 5.0~GHz also found T$_b$ significantly greater than 
10$^{12}$~K (Shen \etal 1997; Tingay \etal 1998). 
The derived core T$_b$ from our 1.6~GHz VSOP image is 
(2.55$\pm$0.66)~$\times$~10$^{12}$~K, which is consistent with 
those earlier estimates. 

It has been shown that there is a limit to T$_b$ for incoherent synchrotron
radiation, and a brightness temperature in excess of this limit is ascribed 
to the effect of Doppler boosting in a relativistic jet beamed toward the 
observer with a Doppler factor 
$\delta$~=~[$\gamma$(1~--~$\beta$cos$\theta$)]$^{-1}$ (cf. Readhead 1994). 
Here $\gamma$~=~(1~--~$\beta^2$)$^{-1/2}$ is the Lorentz factor, $\beta$ is 
the jet velocity in units of the speed of light and, $\theta$ is the angle 
between the line of sight and the radio jet axis.

A commonly accepted explanation is that the observed upper limit to $T_b$
($\sim$~10$^{12}$~K) is caused by the ``inverse Compton catastrophe'' 
(Kellermann, Pauliny-Toth 1969). 
Using formulae (1a) and (1b) rederived by Readhead (1994), we can calculate 
this inverse Compton scattering limit as T$_{b, ic}$~=~1.2~$\times$~10$^{11}$~K 
for PKS~1921--293.
Here we have applied a peak frequency of 8.0~GHz and assumed
a high frequency cutoff of 100~GHz. 
The synchrotron self--absorption turn--over frequency of 8.0~GHz was claimed
by Brown \etal (1989) and is confirmed by single--dish measurements from the 
University of Michigan Radio Astronomy Observatory (UMRAO) made around our 
VSOP observational epoch, from which we also obtained an optically thin spectral 
index of --0.15 as well as a total flux density of 17.9~Jy at 8.0~GHz.
In order to reconcile with T$_{b}$~=~(1.71$\pm$0.44)~$\times$~10$^{12}$~K from 
our 1.6~GHz VSOP results (here, we have multiplied a factor of 0.67 to
convert a brightness temperature derived assuming a Gaussian component to an 
optically thin uniform sphere), a Doppler boosting factor 
$D_{ic}$~=~14.3$\pm$3.7 is required to avoid the inverse Compton catastrophe.
This agrees very well with a lower limit to Doppler factor ($\delta_{ssc}$) of 14
derived from the argument that the observed X-ray emission is produced primarily 
by the inverse Compton scattering of synchrotron radiation (G\"uijosa, Daly 1996
and references therein). 

The inhomogeneous relativistic jet model (Blandford, K\"onigl 1979; 
K\"onigl 1981) also sets an upper limit to the measured brightness temperature. 
This limit is independent of frequency and depends very weakly on the 
observables with an approximate expression as 
T$_{b, j}$~$\sim$~3.0~$\times$~10$^{11}$~$D_{j}$$^{5/6}$~K, here $D_{j}$ is the Doppler
factor associated with this jet model. 
This results in a $D_{j}$~=~12.7$\pm$4.0, which is very similar 
to $D_{ic}$ from the inverse Compton catastrophe.
Combining with the detected superluminal jet motion $\beta_{app}$~=~3.0 (Shen \etal 
1999; choosing $h$~=~0.7 here), we can derive its bulk Lorentz 
factor ($\gamma$) and the jet angle with the line of sight ($\theta$) as follows: 
$\gamma_{ic}$~=~7.5 and 
$\theta_{ic}$~=~1$^\circ$\hspace{-4.5pt}.\hspace{.5pt}6
from $D_{ic}$~=~14.3, and 
$\gamma_i$~=~6.7 and 
$\theta_i$~=~2$^\circ$\hspace{-4.5pt}.\hspace{.5pt}0
from $D_{j}$~=~12.7, respectively. 
Both models require about the same relativistic beaming factor to explain
the high T$_b$ for PKS~1921--293, and we cannot distinguish between them. 

Readhead (1994) introduced the ``equipartition brightness temperature'' cutoff 
($\sim$~10$^{11}$~K) from a statistical analysis. In the case of PKS~1921--293,
it gives a limit of T$_{b, eq}$~=~9.8~$\times$~10$^{10}$~$\delta^{0.78}$~$h^{-2/17}$~K.
The 1.6~GHz VSOP core significantly exceeds this limit, and therefore an
equipartition Doppler factor as large as (39.1$\pm$12.9)~$h^{0.15}$ is needed.
This is about 3 times the values of $D_{ic}$ and $D_{j}$ and suggests that
PKS~1921--293 may not be in equipartition. 
If PKS~1921--293 is not in equipartition, we can use ratio 
$D_{eq}$/$\delta$~=~T$_{b}$/T$_{b, eq}$ with the assumption that $\delta$ is
about 13 (a value close to $D_{ic}$, $D_{j}$ and $\delta_{ssc}$) to derive
an equipartition Doppler factor $D_{eq}$~=~(30.6$\pm$7.9)~$h^{2/17}$, and then
$\gamma_{eq}$~=~14.8 and 
$\theta_{eq}$~=~0$^\circ$\hspace{-4.5pt}.\hspace{.5pt}4.
We can further measure how far the source is from equipartition by calculating 
the ratio $D_{eq}$/$\delta$ which ranges from 2.2~$h^{2/17}$ to 2.4~$h^{2/17}$ 
as $\delta$ changes from 14.3 ($D_{ic}$) to 12.7 ($D_{j}$).
G\"uijosa and Daly (1996) obtained a ratio of 2.0 with $D_{eq}$~=~29 
(assuming $h~=~1$).
This leads to the conclusion that the core of PKS~1921--293 is strongly particle 
dominated, since the particle energy density ($u_p$) greatly exceeds the magnetic 
field energy density ($u_B$) by a factor of ($D_{eq}$/$\delta$)$^{8.5}$, which 
is as large as $\sim$~1000.
Such a departure from equipartition has also been reported for the
superluminal blazar 3C~345 (Unwin \etal 1994). 
We note that both PKS~1921--293 and 3C~345 have not been detected at 
$>$100~MeV $\gamma$--ray energy in spite of the fact that they are among the 
strongest blazars at radio wavelengths. 

Bower and Backer (1998), in their study of the $\gamma$--ray blazar NRAO~530,
favor the inhomogeneous jet model which produces a reasonable Doppler factor
while maintaining energy equipartition.
They also speculate that EGRET--detected blazars are those in which the 
equipartition limit is briefly superseded by the inverse Compton catastrophe limit.
It is also possible that blazars not detected by EGRET may not have equipartition 
between particle and magnetic field energy densities as in the case of PKS~1921--193
while one of the limits imposed by the inverse Compton catastrophe or 
the inhomogeneous relativistic jet model applies.
In any case, a Doppler factor of 12 is required for PKS~1921--293.
This results in a narrow viewing angle 
$\theta$~=~1$^\circ$\hspace{-4.5pt}.\hspace{.5pt}9, at which any 
small bending angle could be enlarged when projected on the sky.
Such phenomenon might be responsible for the observed jet 
curvature and be further related to its non-detection by EGRET 
(cf. Hong \etal 1998; Tingay \etal 1998).

\section{Conclusions}

We have carried out a VSOP observation of the southern blazar PKS~1921--293.
The overall source morphology is consistent with previous ground VLBI 
results. As is clear from figure~2, the space VLBI observations are critical
for isolating the core of the source and permit images to be made with much finer 
spatial resolution than is possible with the ground VLBI at the same frequency. 
In the case of PKS~1921--293,
the high resolution provided by VSOP data, especially along the north-south 
direction, plays an irreplaceable role in our resolving an inner jet component at 
about 1.5~mas north of the compact core. 
When compared with the ground VLBI images, 
this feature is believed to relate to the emission on its curved trajectory from 
the bent jet within 1--2~$h^{-1}$~pc to the 10~pc--scale elongated jets. 

By model fitting VSOP calibrated data, we obtain a core brightness temperature of 
2.6~$\times$~10$^{12}$~K in the source rest frame under the assumption
that the source has a Gaussian brightness distribution. This is in excess
of 10$^{12}$~K, and implies a relativistic beaming in the core. 
We analyzed the source in terms of three models, involving the inverse Compton 
catastrophe, an inhomogeneous relativistic jet, and the equipartition of energy
between the radiating particles and the magnetic field. 
We found no significant difference in Doppler factors for first two models,
though inhomogeneous jet model is more realistic compared to the homogeneous 
sphere model in compact radio sources. 
Both models, however, will eventually lead to a particle dominated 
departure from equipartition state according to the equipartition argument.
Otherwise, a relatively large Doppler factor is needed in order 
to maintain equipartition of energy in the source. 
Thus, our analysis of high $T_b$ in this $\gamma-$ray--quiet blazar PKS~1921--293 
is not in favor of any particular models. 
More VSOP imaging study of these strong blazars with high brightness temperatures 
will be necessary to improve our understanding of the physical process within. \par

\vspace{1pc}\par
We gratefully acknowledge the VSOP Project, which is led by the Japanese 
Institute of Space and Astronautical Science in cooperation with many 
organizations and radio telescopes around the world. This research has made 
use of data from the University of Michigan Radio Astronomy Observatory which 
is supported by the National Science Foundation and by funds from the University 
of Michigan. Research at the ASIAA is funded by the 
Academia Sinica. 

\section*{References} 
\small

\re Blandford R. D., K\"onigl A. 1979, \apj 232, 34
\re Bower G. C., Backer D. C. 1998, \apj 507, L117
\re Brown L. M. J., Robson E. I., Gear W. K., Hughes D. H., Griffin M. J.,
    Geldzahler B. J., Schwartz P. R., Smith M. G. \etal 1989, \apj 340, 129
\re de Pater I., Schloerb F. P., Johnson A. H. 1985, \aj 90, 846
\re G\"uijosa A., Daly R. A. 1996, \apj 461, 600
\re Fey A. L., Clegg A. W., Fomalont E. B. 1996, \apjs 105, 299
\re Fichtel C. E., Bertsch D. L., Chiang J., Dingus B. L., Esposito J. A., Fierro J. M.,
    Hartman R. C., Hunter S. D. \etal 1994, \apjs 94, 551
\re Hirabayashi H., Hirosawa H., Kobayashi H., Murata Y., Edwards P. G., Fomalont E. B.,
    Fujisawa K., Ichikawa T. \etal 1998, \sci 281, 1825 and erratum 282, 1995
\re Hong X. Y., Jiang D. R., Shen Z.-Q. 1998, \aap 330, L45
\re Kellermann K. I., Pauliny-Toth I. I. K. 1969, \apj 193, 43
\re Kellermann K. I., Vermeulen R. C., Zensus J. A., Cohen M. H. 1998,
	   \aj, 115, 1295
\re K\"onigl A. 1981, \apj 243, 700
\re Linfield R. P., Levy G. S., Ulvestad J. S., Edwards C. D., DiNardo S. J., 
    Stavert L. R., Ottenhoff C. H., Whitney A. R. \etal 1989, \apj 336, 1105
\re Moellenbrock G. A., Fujisawa K., Preston R. A., Gurvits L. I., Dewey R. J.,
    Hirabayashi H., Inoue M., Kameno S. \etal 1996, \aj 111, 2174 
\re Mukherjee R., Bertsch D. L., Bloom S. D., Dingus B. L., Esposito J. A., Fichtel C. E.,
    Hartman R. C., Hunter S. D. \etal 1997, \apj 490, 116
\re Perley R. A. 1982, \aj 87, 859 
\re Readhead A. C. S. 1994, \apj 426, 51
\re Shen Z.-Q., Moran J.M., Kellermann K.I. 1999, in preparation
\re Shen Z.-Q., Wan T.-S., Moran J. M., Jauncey D. L., Reynolds J. E., Tzioumis A. K.,
    Gough R. G., Ferris R. H. \etal 1997, \aj 114, 1999
\re Shepherd M. C. 1997, in Astronomical Data Analysis Software and Systems VI, 
           ASP Conf. Series 125, eds. G. Hunt \& H. E. Payne (San Francisco: ASP), 77
\re Tingay S. J., Murphy D. W., Lovell J. E. J., Costa M. E., McCulloch P., Edwards P. G.,
    Jauncey D. L., Reynolds J. E. \etal 1998, \apj 497, 594 
\re Tingay S. J., Murphy D. W., Edwards P. G. 1998, \apj 500, 673
\re Unwin S. C., Wehrle A. E., Urry C. M., Gilmore D. M., Barton E. J., Kjerulf B. C.,
    Zensus J. A., Rabaca C. R. 1994, \apj 432, 103
\re Wills D., Wills B. J. 1981, \nat 289, 384
\re Worrall D. M., Wilkes B. J. 1990, \apj 360, 396

\label{last}

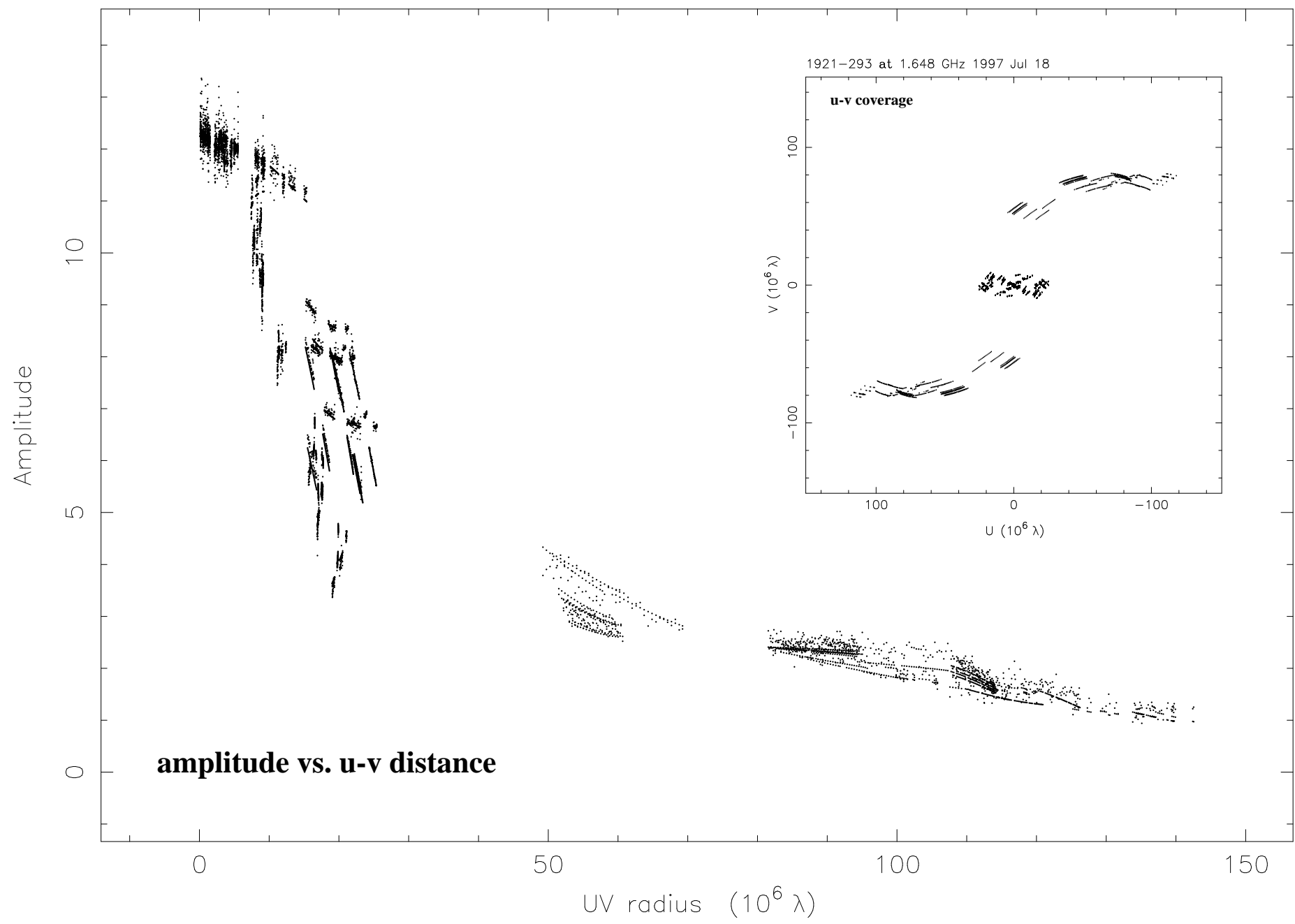
\end{document}